\documentstyle[12pt]{article}
\begin{document}

\newcommand{\uuu}{\mbox{\huge $u$}}
\newcommand{\vvv}{\mbox{\huge $v$}}
\newcommand{\xxx}{\end{document}}
\newcommand{\cue}{{\cal E}}
\newcommand{\cun}{{\mbox{\scriptsize${\cal N}$}}}
\newcommand{\bcc}{\begin{center}}
\newcommand{\ecc}{\end{center}}
\def \inbar{\vrule height1.5ex width.4pt depth0pt}
\def \C{\relax\hbox{\kern.25em$\inbar\kern-.3em{\rm C}$}}
\def \R{\relax{\rm I\kern-.18em R}}
\newcommand{\Z}{\ Z \hspace{-.08in}Z}
\newcommand{\be}{\begin{equation}}
\newcommand{\ee}{\end{equation}}
\newcommand{\bea}{\begin{eqnarray}}
\newcommand{\eea}{\end{eqnarray}}
\newcommand{\p}{\psi}
\newcommand{\f}{\phi}
\newcommand{\g}{\gamma}
\newcommand{\G}{\Gamma}
\newcommand{\e}{\eta}
\newcommand{\m}{\mu}
\newcommand{\n}{\nu}
\newcommand{\s}{\sigma}
\renewcommand{\t}{\tau}
\renewcommand{\d}{\delta}
\renewcommand{\pt}{\frac{\partial}{\partial t}}
\newcommand{\ppt}{\frac{\partial^{2}}{\partial t^{2}}}
\newcommand{\nn}{\nonumber}
\newcommand{\kt}{\rangle}
\newcommand{\br}{\langle}
\newcommand{\fs}{\small}
\newcommand{\so}{S_{0}}
\newcommand{\I}{\mbox{1}_{m\times m}}
\newcommand{\In}{\mbox{1}_{n\times n}}
\newcommand{\xo}{x_{0}}
\newcommand{\po}{\psi_{0}}
\newcommand{\eo}{\e_{0}}
\newcommand{\ts}{\tilde{S}}
\newcommand{\pss}{\frac{\partial}{\partial s}}
\newcommand{\tcuf}{\tilde{\cal F}}
\newcommand{\cuf}{{\cal F}}
\newcommand{\oot}{\mbox{\fs$\frac{1}{2}$}}
\newcommand{\iot}{\mbox{\fs$\frac{i}{2}$}}
\newcommand{\cur}{{\cal R}}
\newcommand{\iv}{\imath\! v}
\newcommand{\ib}{\int_{0}^{\b}}
\newcommand{\lll}{\left( }
\newcommand{\rrr}{\right)}
\newcommand{\llc}{\left\{ }
\newcommand{\rrc}{\right\} }
\newcommand{\lpt}{\left.}
\newcommand{\rpt}{\right.}
\newcommand{\rar}{\longrightarrow}
\newcommand{\lar}{\longleftarrow}
\newcommand{\cinf}{C^{\infty}\!}
\newcommand{\vol}{d\mbox{\bf\fs$\Omega$}}
\newcommand{\sx}{\mbox{\fs$(x)$}}
\newcommand{\hD}{\hat{D}}
\newcommand{\V}{(V)}
\newcommand{\La}{\Lambda}
\newcommand{\ph}{{\cal P}({\cal H})}
\newcommand{\cp}{\C\! P}

\newcommand{\qq}{{\cal Q}}
\newcommand{\psqm}{$(p=2)$--PSQM~ }
\newcommand{\eeqq}{|E,q_1\kt}
\newcommand{\qpee}{\br E,q_1'|}
\newcommand{\qqee}{\br E,q_1|}

\title{Spectrum Degeneracy of General $(p=2)$--Parasupersymmetric Quantum
Mechanics and Parasupersymmetric Topological Invariants}
\author{\\ Ali Mostafazadeh\thanks{Supported by Killam Postdoctoral
Fellowship.}\\  ~ \\Theoretical Physics Institute, Department of Physics,\\
University of Alberta, Edmonton, AB, Canada T6G 2J1}
\date{~}
\maketitle
\abstract{ A thorough analysis of the general features of $(p=2)$
parasupersymmetric quantum mechanics is presented.  It is shown that for both
Rubakov--Spiridonov and Beckers--Debergh formulations of
$(p=2)$-parasupersymmetric quantum mechanics,  the degeneracy structure of the
energy spectrum can be derived using   the defining parasuperalgebras.  Thus
the results of the present article is independent of the details of the
Hamiltonian. In fact, they are valid for arbitrary systems based on  arbitrary
dimensional coordinate manifolds.  In particular,  the
Rubakov--Spiridonov (R-S) and  Beckers--Debergh (B-D) systems possess identical
degeneracy  structures. For a subclass of R-S (alternatively B-D) systems,
a new topological invariant is introduced. This is a counterpart of the Witten
index of the supersymmetric quantum mechanics. }


\section{Introduction}
There are two alternative definitions of $(p=2)$--parasupersymmetric quantum
mechanics (PSQM). These are the original Rubakov--Spiridonov \cite{rs}, and the
Beckers--Debergh \cite{bd} formulations of $(p=2)$--PSQM. The
Rubakov--Spiridonov (R-S) $(p=2)$--PSQM is defined by the parasuperalgebra:
	\bea
	\qq^3&=&0 	\label{e1}\\
	\left[ H,\qq\right]&=&0	\label{e2}\\
	\qq^2\qq^\dagger+\qq\qq^\dagger\qq+\qq^\dagger\qq^2&=&4\qq H \;, \label{e3}
	\eea
where $\qq$ is a parasupercharge and $H$ is the Hamiltonian. The R-S \psqm ~has
been studied \cite{rs} and its variations and generalizations to arbitrary
orders ($p>2$) have been given  by Khare \cite{kh}.  The degeneracy structure
of this type of PSQM has been worked out  for specific examples and some
classes of systems in ordinary one-dimensional quantum mechanics, \cite{rs,kh}.

Similarly, the Beckers--Debergh (B-D) \psqm is defined by the parasuperalgebra:
	\bea
	\qq^3&=&0	\label{e4}\\
	\left[ H,\qq\right]&=&0	\label{e5}\\
	\left[\qq ,\left[\qq^\dagger,\qq\right]\right]&=&2\qq H\; .
	\label{e6}
	\eea
Particular examples of this  type of PSQM has  been studied \cite{bd} in the
context of one-dimensional quantum mechanics. The corresponding coherent states
 have also been constructed \cite{bd2}.

In this article the parasuperalgebras (\ref{e1})--(\ref{e3}) and
(\ref{e4})--(\ref{e6}) are used to study the degeneracy structure of the most
general R-S and B-D systems. Our strategy is analogous to the one used for the
treatment of supersymmetric quantum mechanics (SQM), \cite{susy,witten}.

One of the most intriguing aspects of SQM is its relation to the Atiyah--Singer
index theorem
\cite{witten} (For an up-dated account of this subject see \cite{mo} and the
references therein.)
As SQM can be viewed as the $(p=1)$--PSQM, one might be tempted to seek similar
features of
($p=2$) or even ($p>2$)--PSQM. The first step in this direction is to explore
the degeneracy structure of such systems. This is the main motivation behind
our general treatment of \psqm .

Before pursuing the study of \psqm, we present a brief review of the relevant
aspects of SQM in Sec.~2. The R-S and B-D \psqm are analyzed in Secs.~3 and~4,
respectively. In Sec.~5, a subclass of R-S (alternatively B-D) \psqm
is considered. For this class a topological invariant is introduced which
resembles the Witten index of SQM. Sec.~6 includes the concluding remarks.


\section{Supersymmetric Quantum\ Mechanics}
The $(N=1)$-SQM is defined according to the superalgebra:
	\bea
	\{ \qq ,\qq \} & =& [H,\qq ]\: =\: 0	\label{a1} \\
	\{\qq,\qq^\dagger\}&=&2\kappa H\; ,\label{a2}
	\eea
where  $\qq$ and $H$ stand for the supercharge and the Hamiltonian and $\kappa$
is a
positive real number. It is usually chosen to be  1. However, this convention
does not agree with the conventions used in different  approaches to \psqm.

 Furthermore, there
exist a self-adjoint involution $\mbox{\large$\tau$}$  on the Hilbert space
${\cal H}$ which satisfies:
	\bea
	\mbox{\large$\tau$}^2&=&1,  \label{t1}\\
	\{\mbox{\large$\tau$},\qq\} &=&[\mbox{\large$\tau$},H]\:=\:0\;.  \label{t2}
	\eea
The involution $\mbox{\large$\tau$}$  introduces a double grading of the
Hilbert space, i.e., it leads to a decomposition of  the  Hilbert space ${\cal
H}$ into its  $\pm 1$--eigenspaces:
	\[ {\cal H}={\cal H}_{-}\oplus{\cal H}_+\:,~~~{\rm with}~~~\mbox{\large$\tau$}
|\psi_{\pm}\kt 	=\pm|\psi_\pm\kt\;,~~~ \forall |\psi_\pm\kt\in{\cal H}_\pm\;.\]
The involution $\mbox{\large$\tau$}$ is also called the chirality operator and
denoted by $(-1)^F$.
The $\pm 1$--eigenstates of $\mbox{\large$\tau$}$ are said to be even (for +)
and odd (for --) elements of the Hilbert space.  They are also said to have
$\pm$ chirality. For more details of the importance of
the chirality operator see \cite{chirality}.

An important property of SQM is that its degeneracy structure can be easily
obtained  using  the superalgebra  (\ref{a1})--(\ref{a2}) and the properties of
chirality operator (\ref{t1}) and (\ref{t2}).
This is best carried out by introducing   the self-adjoint supercharges:
	\be
Q_1:=\frac{1}{\sqrt{2}}(\qq+\qq^\dagger)\;,~~Q_2:=
\frac{-i}{\sqrt{2}}(\qq-\qq^\dagger)\;,
	\label{e7}
	\ee
and expressing Eqs.~(\ref{a1}),~(\ref{a2}), and (\ref{t2}) in terms of them.
The result is
	\bea
	\{Q_1,Q_2\}&=&[H,Q_1]\: =\: [H,Q_2]\;=\:0	\label{aa1}\\
	H&=&\kappa^{-1}Q_1^2\:=\:\kappa^{-1}Q_2^2 \label{aa2}\\
	\{\mbox{\large$\tau$},Q_1\}&=&\{\mbox{\large$\tau$},Q_2\}\:=\:
[\mbox{\large$\tau$}, H]\:=\:0\; .\label{tt2}
	\eea

To start the analysis of the spectrum of SQM, one uses the eigenvalues of $H$
an (say)
$Q_1$ to label the basic states $|E,q_1\kt$, where
	\bea
	H|E,q_1\kt&=&E|E,q_1\kt	\label{e13}\\
	Q_1\eeqq&=&q_1\eeqq\;. 	\label{e14}
	\eea
Combining Eqs.~(\ref{e13}),(\ref{e14}), and (\ref{aa2}), one immediately finds:
	\[ E\eeqq=H\eeqq=\kappa^{-1}Q_1^2\eeqq=\kappa^{-1}q_1^2\eeqq\;,\]
which for $\eeqq\neq 0$ implies $q_1=\pm\sqrt{\kappa E}$. Thus
\begin{itemize}
\item[a)] $E\geq 0$ ;
\item[b)] $E=0$ is non-degenerate;
\item[c)] $E>0$ are doubly degenerate.
\end{itemize}
Note that in practice there may exist other conserved quantities (observables)
that would
introduce further quantum numbers and thus correspond to additional
degeneracies associated with  each $\eeqq$. We shall call these
subdegeneracies.  The existence of this conserved quantities  overshadows the
effectiveness of statement  (b) above, but  statement (c)  is still worth
investigating.

In addition, Eqs.~(\ref{aa1}) and (\ref{tt2}) can be easily employed to show
	\[ Q_2|E, \pm\sqrt{\kappa E}\kt=\sqrt{\kappa E}e^{\pm i\phi} \,
	|E,\mp\sqrt{\kappa E}\kt\;.\]
In fact, the phase factors $e^{\pm i\phi}$ can be absorbed in the definition of
  $|E, \pm\sqrt{\kappa E}\kt$, so that
	\be
	Q_2|E, \pm\sqrt{\kappa E}\kt=\sqrt{\kappa E}|E,\mp\sqrt{\kappa E}\kt\;.
	\label{x1}
	\ee
Since $Q_2$ anticommutes with $\mbox{\large$\tau$}$  (it is   an odd operator),
 $|E, \pm\sqrt{\kappa E}\kt$ are called superpartners. To motivate the use of
the word superpartner, one can construct a basis of the $E>0$
subspaces consisting of a bosonic (even) and a fermionic (odd) state vector.
In the basis
	$$\left\{ |E, \sqrt{\kappa E}\kt=\lll \begin{array}{c}1\\0\end{array}\rrr,
	 |E,-\sqrt{\kappa E}\kt=\lll \begin{array}{c}0\\1 \end{array}\rrr \right\}$$
one has:
	\be
	\left. Q_1\right|_{{\cal H}_E}=\sqrt{\kappa E} \lll
	\begin{array}{cc}
	1&0\\
	0&-1
	\end{array} \rrr=\sqrt{\kappa E}\mbox{\large$\sigma$}_3~,~~~~~
	\left. Q_2\right|_{{\cal H}_E}=\sqrt{\kappa E}\lll
	\begin{array}{cc}
	0&1\\
	1&0
	\end{array}\rrr=\sqrt{\kappa E}\mbox{\large$\sigma$}_1 \;.
	\label{x2}
	\ee
Here ${\cal H}_E$ denotes the degeneracy subspace associated with the energy
level $E$ (Note that $E>0$). Using Eqs.~(\ref{t1}) and (\ref{tt2}) and the the
fact that the chirality operator is self-adjoint, one can show that there are
only two possibilities for $\mbox{\large$\tau$}$. These are
	\be
	\left. \mbox{\large$\tau$} \right|_{{\cal H}_E}=\eta \lll\begin{array}{cc}
	0&-i\\
	i&0
	\end{array}\rrr=\eta\mbox{\large$\sigma$}_2\;,
	\label{x3}
	\ee
where in Eqs.~(\ref{x2}) and (\ref{x3}), $\mbox{\large$\sigma$}_a$ with
$a=1,2,3$ are Pauli matrices, and $\eta=\pm 1$. The arbitrariness of the sign
proves to be unimportant. An exchange of sign corresponds to an exchange of the
eigenvalues ($\pm 1$) of $\mbox{\large$\tau$}$. In any case, det$(\left.
\mbox{\large$\tau$}\right|_{{\cal H}_E})=-1$.

Having found the matrix representation of $\mbox{\large$\tau$}$, one can
diagonalize it and find a
basis of the $E$-eigenspace consisting of the state vectors of definite
chirality. These are
	\bea
	|E,+\kt &=&\frac{1}{\sqrt{2}}(|E,\sqrt{\kappa E}\kt+i|E,-\sqrt{\kappa E}\kt)
\label{x4}\\
	| E,-\kt &=&\frac{1}{\sqrt{2}}(|E,\sqrt{\kappa E}\kt-i|E,-\sqrt{\kappa
E}\kt)\;, \label{x5}
	\eea
with $\mbox{\large$\tau$} |E,\pm\kt=\pm|E,\pm\kt$. Choosing the opposite sign
for $\eta$ (respectively for $\mbox{\large$\tau$}$) leads to changing
$|E,\pm\kt\to|E,\mp\kt$ in (\ref{x4}) and (\ref{x5}).

Therefore, in general each ($E>0$)--level consists of two linearly independent
vectors of opposite chirality.
These are the true superpartners.  Even if there are further subdegeneracies,
(in view of  the statement (c) above) the positive energy levels  must include
pairs of superpartners.  Thus there must be equal number of bosonic and
fermionic states with positive energy. This is
the very reason why the Witten index  (=trace$(\mbox{\large$\tau$})$) is
invariant under smooth deformations of the Hamiltonian \cite{witten} and how
SQM is related to the topological invariants such as indices of elliptic
operators \cite{mo}. A simple manifestation of the properties of the Witten
index is demonstrated by (\ref{x3}) which clearly indicates
trace($\left.\mbox{\large$\tau$}\right|_{{\cal H}_{E>0}}$)=0.

The main purpose of the present article is to follow a similar approach in the
study  of  \psqm.


\section{Rubakov--Spiridonov \psqm}
Let us first express the R-S parasuperalgebra (\ref{e1})--(\ref{e3}) in terms
of the self-adjoint parasupercharges (\ref{e7}):
	\bea
	&&Q_1^3-\{Q_1,Q_2^2\}-Q_2Q_1Q_2=0	\label{e8}\\
	&&Q_2^3-\{Q_2,Q_1^2\}-Q_1Q_2Q_1=0
	\label{e9}\\
	&&\left[H,Q_1\right]\:=\:\left[H,Q_2\right]=0	\label{e10}\\
	&&Q_1^3\:=\:2Q_1H~		\label{e11}\\
	&&Q_2^3\:=\:2Q_2H\;,		\label{e12}
	\eea
where (\ref{e8}) and (\ref{e9}) are equivalent to (\ref{e1}), (\ref{e10}) is
equivalent to (\ref{e2}), and (\ref{e11}) and (\ref{e12}) are equivalent to
(\ref{e3}). Clearly, Eq.~(\ref{aa2}) with $\kappa=2$ is a special case of
Eqs.~(\ref{e8})--(\ref{e12}). Thus, SQM is included in R-S \psqm.

Combining Eqs.~(\ref{e8}) and (\ref{e11}) and similarly Eqs.~(\ref{e9}) and
(\ref{e12}), one obtains the following useful relations:
	\bea
	2Q_1H-\{Q_1,Q_2^2\}-Q_2Q_1Q_2&=&0	\label{e1.7}\\
	2Q_2H-\{Q_2,Q_1^2\}-Q_1Q_2Q_1&=&0\;.
	\label{e1.8}
	\eea

Again one can appeal to Eq.~(\ref{e10}) to choose $\eeqq$ of Eqs.~(\ref{e13})
and (\ref{e14}) as basic state vectors.  In view of Eqs.~(\ref{e11}),
(\ref{e13}) and (\ref{e14}), one has:
	\[ (Q_1^3-2Q_1H)\eeqq=q_1(q_1^2-2H)\eeqq\;.\]
Thus,
	\be
	{\rm either}~~q_1=0\;,~~~~~{\rm or}~~q_1=\pm\sqrt{2E}\:.  \label{e16}
	\ee
In particular if $E\leq 0$, one necessarily has $q_1=0$. Thus the non-positive
energy levels -- if they exist -- are non-degenerate.\footnote{That is aside
from the subdegeneracies.}

To study the positive energy levels, first we write Eqs.~(\ref{e1.7}) and
(\ref{e1.8}) in terms of their matrix elements:
	\bea
	\qpee\left[ 2q_1E-(q_1+q_1')Q_2^2-Q_2Q_1Q_2\right]\eeqq&=&0 \label{e1.9}\\
	\left[ 2E-(q_1+q_1')^2+q_1q_1'\right]\qpee Q_2\eeqq&=&0\;. \label{e1.10}
	\eea
Next we analyze the following possibilities:
	\begin{itemize}
	\item[~]{\bf Case (1)}: $E>0$ and $|E,q_1=0\kt\neq 0$;
	\item[~]{\bf Case (2)}: $E>0$ and $|E,q_1=0\kt = 0$.
	\end{itemize}

Consider Case (1). Then one can set $q_1=q_1'=0$ in Eq.~(\ref{e1.10}) and
obtain:
	\be
	\br E,0|Q_2|E,0\kt=0\;, \label{e2.1}
	\ee
Therefore, either $E$ is non-degenerate and $Q_2|E,0\kt=0$, or it is degenerate
and there is some $q_1\neq 0$ such that $|E,q_1\kt\neq 0$. We shall refer to
these two cases as Case~(1a) and Case~(1b), respectively. Next we consider
Case~(1b).

In this case, Eq.~(\ref{e1.10}) together with (\ref{e16}) lead to:
	\be
	\qqee Q_2\eeqq=0\;. 	\label{e3.1}
	\ee
Note that, {\em a priori}, $q_1$ may take one or both possible values
$\pm\sqrt{2E}$. In other words, there are again two possibilities:
	\begin{itemize}
	\item[I)]  $|E,\pm\sqrt{2E}\kt\neq 0$
	\item[II)] $|E,q_1\kt\neq 0$ and $|E,-q_1\kt=0$, for $q_1=\sqrt{2E}$,
	or $q_1=-\sqrt{2E}$.
	\end{itemize}

In general, according to Eq.~(\ref{e2.1}):
	\be
	Q_2|E,0\kt = a_+|E,\sqrt{2E}\kt+a_-|E,-\sqrt{2E}\kt\;.
	\label{e18}
	\ee
where $a_\pm$  are complex coefficients. Moreover, setting $q_1=q_1'=0$ in
Eq.~(\ref{e1.9}), one finds:
	\be
	\br E,0|Q_2Q_1Q_2|E,0\kt=0\;.	\label{e19}
	\ee
This relation and Eq.~(\ref{e18}) can be used to show
	\bea
	a_+^*a_+&=&a_-^*a_-\;,~~~{\rm if} ~~|E,\pm\sqrt{2E}\kt\neq 0\;;
	\label{e2.3}\\
	a_\pm&=&0\;,~~~~{\rm if} ~~|E,\sqrt{2E}\kt=0~~{\rm or}~~
	|E,-\sqrt{2E}\kt=0\;.	\label{e2.4}
	\eea

Let us show that in fact  Case (II) above does not occur, i.e.,  the vanishing
of
either of $|E,\pm\sqrt{2E}\kt$ implies the vanishing of the other.
By contradiction assume $|E,\sqrt{2E}\kt\neq 0$ and $|E,-\sqrt{2E}\kt=0$.
Then according to (\ref{e2.4}), $a_\pm=0$ and $Q_2|E,0\kt=0$.  This together
with Eq.~(\ref{e3.1}) imply,  $Q_2|E,\sqrt{2E}\kt \propto |E,0\kt$.  In fact,
since $Q_2|E,0\kt$ vanishes, $Q_2^2|E,\sqrt{2E}\kt=0$. Hence
$Q_2|E,\sqrt{2E}\kt=0$ as well.
Using the last equation and Eqs.~(\ref{e8}) and (\ref{e16}), one is led to the
result
$E=0$. A similar argument shows that $|E,\sqrt{2E}\kt=0$ and
$|E,-\sqrt{2E}\kt\neq 0$  give rise to the same conclusion. This  contradicts
the hypothesis ($E>0$). Thus this case is forbidden and the energy levels of
type (1b) are triply degenerate.

To obtain the matrix representation of  $Q_2$ in Case~(1b), one rewrites
Eq~(\ref{e3.1}) in the form:
	\be
	\begin{array}{c}
	Q_2|E,\sqrt{2E}\kt=c_+|E,0\kt+f|E,-\sqrt{2E}\kt\\
	Q_2|E,-\sqrt{2E}\kt=c_-|E,0\kt+f ^*|E,\sqrt{2E}\kt\;.
	\end{array}
	\label{w0}
	\ee
The coefficients $c_\pm$ and  $a_\pm$  can be related by computing
$|Q_2|E,q_1\kt|^2$ and $\br E,q_1|\lll Q_2^2|E,q_1\kt\rrr$ and equating the
results. This leads to
	\bea
	c_+(c_+^*-a_+)&=&0	\label{w1}\\
	c_-(c_-^*-a_-)&=&0	\label{w2}\\
	|a_\pm|^2&=&\frac{1}{2}(a_+c_++a_-c_-)\;,	\label{w3}
	\eea
Furthermore, setting $q_1=q_1'=\pm\sqrt{2E}$ in (\ref{e1.9}) and using
(\ref{w0}), one finds:
	\be
	|f|^2=2(E-|c_\pm|^2)\;.
	\label{w4-5}
	\ee
Hence, $|c_+|=|c_-|$. A final relation among $a_\pm,~c_\pm$ and $f$ is obtained
by acting both sides of (\ref{e12}) on $|E,0\kt$ and repeatedly using
(\ref{e18}) and (\ref{w0}). This yields:
	\be
	c_+c_-^*f^*+c_+^*c_-f=0\;.
	\label{w10}
	\ee

In view of Eqs.~(\ref{w1}), (\ref{w2}), and (\ref{w3}), there are two possible
cases:
	\begin{itemize}
	\item[ ~]{\bf Case~(1.b.i)}:~ $a_\pm=c_\pm^*\neq 0$, in which case
$Q_2|E,0\kt\neq 0$;
	\item[ ~]{\bf Case~(1.b.ii)}:~  $a\pm=c_\pm=0$, in which case $Q_2|E,0\kt= 0$.
	\end{itemize}

For convenience we introduce the real parameter $\zeta\in [0,1]$:
	\be
	\zeta:=\frac{|a_\pm|}{\sqrt{E}}=\frac{|c_\pm|}{\sqrt{E}}\;.
	\label{zeta}
	\ee
Then one has:
	\[a_\pm=\zeta\sqrt{E}  e^{-i\gamma_\pm}\;,c_\pm=\zeta\sqrt{E}
e^{i\gamma_\pm}\;,
	f=\sqrt{2E(1-\zeta^2)}e^{i\varphi}\;.\]
In view of these relations and Eq.~(\ref{w10}), (\ref{e18}) and (\ref{w0}) take
the form:
	{\small \bea
	Q_2|E,0\kt&=&\zeta\sqrt{E}\lll e^{-i\gamma_+}|E,\sqrt{2E}\kt
+e^{-i\gamma_-}|E,-\sqrt{2E}\kt\rrr
	\label{e18'} \\
	Q_2|E,\pm\sqrt{2E}\kt&=&\!\!\!\sqrt{E}\left[ \zeta
e^{i\gamma\pm}|E,0\kt+\sqrt{2(1-\zeta^2)}
	e^{\pm i\varphi}|E,\mp\sqrt{2E}\kt\right]~~~~~~
	\label{w0'}\\
	e^{i(\varphi-\gamma_++\gamma_-)}&=&i\epsilon,~~~{\rm with~~~}\epsilon=\pm\;.
	\label{w10'}
	\eea}
Redefining $|E,\pm\sqrt{2E}\kt\to e^{-i\gamma_\pm}|E,\pm\sqrt{2E}\kt$, one can
eliminate  all the phase factors. This yields
	\bea
	Q_2|E,0\kt&=&\zeta\sqrt{E}\lll |E,\sqrt{2E}\kt
	+|E,-\sqrt{2E}\kt\rrr
	\label{e18''}\\
	Q_2|E,\pm\sqrt{2E}\kt&=&\sqrt{E}\lll\zeta |E,0\kt\pm i\epsilon
\sqrt{2(1-\zeta^2)}
	|E,\mp\sqrt{2E}\kt\rrr \;.
	\label{w0''}
	\eea

Remarkably, the defining relations (\ref{e8})--(\ref{e12}) of R-S \psqm, do not
impose any restriction on $\zeta$ and $\epsilon$. Hence, unless further details
of the system  is known,
they cannot be determined.

At this stage, we can try to find a representation of the parasupercharges.
Using the basis
	\[ \left\{ |E,\sqrt{2E}\kt=\lll \begin{array}{c}
	1\\0\\0\end{array}\rrr~,~|E,0\kt=\lll\begin{array}{c}
	0\\1\\0\end{array}\rrr~,~|E,-\sqrt{2E}\kt=\lll\begin{array}{c}
	0\\0\\1\end{array}\rrr\right\}\;,\]
one can easily express $Q_1$ and $Q_2$ as $3\times 3$ matrices:
	\be
	\begin{array}{c}
	\left. Q_1\right|_{{\cal H}_E}\: =\: \sqrt{2E}\lll
	\begin{array}{ccc}
	1&0&0\\
	0&0&0\\
	0&0&-1
	\end{array}\rrr=\sqrt{2E}J_3^{~(1)} \;,    \\ \\
	\left. Q_2\right|_{{\cal H}_E}\: =\: \sqrt{2E}\lll
	\begin{array}{ccc}
	0&	\frac{\zeta}{\sqrt{2}}    &-i\epsilon\sqrt{1-\zeta^2}\\
	\frac{\zeta}{\sqrt{2}}&0&\frac{\zeta}{\sqrt{2}}\\
	i\epsilon\sqrt{1-\zeta^2}&\frac{\zeta}{\sqrt{2}}&0
	\end{array}\rrr \\ \\
	\hspace{1.5cm}	=\sqrt{E}\zeta J_1^{~(1)}+i\epsilon\sqrt{2E(1-\zeta^2)}\lll
	\begin{array}{ccc}
	0&0&-1\\
	0&0&0\\
	1&0&0
	\end{array}\rrr\;.
	\end{array}
	\label{rep}
	\ee
where $J_i^{~(1)}$, with $i=1,2,3$, are the three dimensional ($j=1$)
representation of the generators of $SU(2)$.  One may also check that for all
values of $\zeta$ and $\epsilon$,
${\rm Spectrum}(\left. Q_2\right|_{{\cal H}_E})=\{0,\pm\sqrt{2E}\}={\rm
Spectrum}(\left. Q_1\right|_{{\cal H}_E})$.

Next step is to adopt a self-adjoint involution operator $\mbox{\large$\tau$}$
which satisfies (\ref{t1}) and (\ref{tt2}) and use these equations to find its
representation in the $E$-eigenspace. Using the self-adjointness of
$\mbox{\large$\tau$}$ and imposing Eqs.~(\ref{t1}) and (\ref{tt2}), one finds
	\be
	\left. \mbox{\large$\tau$}\right|_{{\cal H}_E}=\lll\begin{array}{ccc}
	0&0&\tilde\eta\\
	0&\eta&0\\
	\tilde\eta&0&0
	\end{array}\rrr\;,
	\label{tau}
	\ee
where, for $\zeta=0$ (Case~(1.b.ii)~), $\eta$ and $\tilde\eta$ are arbitrary
signs, i.e., $\eta,\tilde\eta =\pm 1$, and for $\zeta\neq 0$ (Case~(1b.i)~),
$\tilde\eta=-\eta=\pm 1$. The sign ambiguities in (\ref{tau}) is in a sense
analogous to the case of SQM. However as we show below,  there are some
important  differences.

Let us construct the states of definite chirality. These are
	\be
	|E,\pm\kt:=\frac{1}{\sqrt{2}}\lll \begin{array}{c}
	1\\0\\ \pm\tilde\eta \end{array}\rrr\;~~~~|E,\eta^\circ\kt:=\lll
	\begin{array}{c}
	0\\1\\0\end{array}\rrr\;.
	\label{tau1}
	\ee
They satisfy:
	\be
\mbox{\large$\tau$}|E,\pm\kt=\pm|E,\pm\kt\;,~~~
\mbox{\large$\tau$}|E,\eta^\circ\kt=
	\eta|E,\eta^\circ\kt\;.\label{tau2}
	\ee
In other words  depending on the value of $\eta=\pm1$,
$|E,\eta^\circ=\pm^\circ\kt$ is parabosonic or parafermionic.  Note also that
$\eta=-det(\left. \mbox{\large$\tau$}\right|_{{\cal H}_E})$. Thus choosing
$det(\left. \mbox{\large$\tau$}\right|_{{\cal H}_E})=-1$, as is the case in
SQM,  is equivalent to  setting  $\eta=1$.

Next we display the action of $Q_2$ on the states of definite chirality:
	\bea
	Q_2|E,\pm\kt&=&\sqrt{2E}\left[
	\mp i\epsilon\tilde\eta\sqrt{1-\zeta^2}~|E,\mp\kt +
	\frac{1\pm\tilde\eta}{2}\zeta~|E,\eta^\circ\kt\right]
	\label{tau3}\\
	Q_2|E,\eta^\circ\kt&=&\zeta \sqrt{E}|E,-\eta\kt \;.
	\label{tau4}
	\eea
There are two specially interesting cases. These are characterized by $\zeta=0$
(Case~(1.b.ii)~), for which
	\be
	Q_2|E,\pm\kt=\mp
i\epsilon\tilde\eta\sqrt{2E}|E,\mp\kt\;,~~~Q_2|E,\eta^\circ\kt=0\;,
	\label{tau5}
	\ee
and $\zeta=1$ for which
	\be
	\begin{array}{c}
	Q_2|E,\pm\kt=\sqrt{2E}(\frac{1\mp\eta}{2})|E,\eta^\circ\kt\;,~~~
	Q_2|E,\eta^\circ\kt=\sqrt{E}|E,-\eta\kt\\
	\left. Q_2\right|_{{\cal H}_E}=\sqrt{2E}\lll\begin{array}{ccc}
	0&\frac{1}{\sqrt{2}}&0\\
	\frac{1}{\sqrt{2}}&0&\frac{1}{\sqrt{2}}\\
  	0&\frac{1}{\sqrt{2}}&0
	\end{array}\rrr=\sqrt{2E}J_1^{~(1)}\:.
	\end{array}
	\label{tau6}
	\ee
For these two special cases $Q_2$ eliminates one of the states defined in
(\ref{tau1}). Moreover,
as we shall see in Sec.~4,  these two cases  also appear  in the B-D \psqm.
The coincidence of $Q_1$ and $Q_2$ with the generators of $SU(2)$ is analogous
with SQM.

One must emphasize that  unlike SQM, in \psqm different choices of {\large
$\tau$} (different choices for
$\eta$ and $\tilde\eta$) can lead to different systems. This is because here
the energy levels do not consist of pairs of superpartners. In fact, in general
the defining signs $\eta$ and $\tilde\eta$ which appear in the expression
(\ref{tau}) of  {\large$\left. \tau\right|_{{\cal H}_E}$} depend on $E$. They
may take different values for different energy levels. Thus as far as
quantities such as
 the difference of the numbers of parabosonic and parafermionic  states are
concerned, the values of $\eta=\eta(E)$ and $\tilde\eta=\tilde\eta(E)$ are
important. For $\zeta\neq 0$, requiring $det( \mbox{\large$\left.
\tau\right|_{{\cal H}_E}$})=-1$ for all $E$, fixes $\eta(E)=-\tilde\eta(E)=1$.
In this case, the energy levels of type~(1b) consist of two parabosonic states
and one parafermionic state.

This completes our analysis of Case~(1).  A summary of the results  is in
order:
\newtheorem{lem}{Lemma}
\begin{lem}
 For $E>0$ and $|E,0\kt\neq 0$, either $E$ is non-degenerate with
$Q_2|E,0\kt=0$ or it is triply degenerate with the basis $\{
|E,0\kt,|E,\pm\sqrt{2E}\kt\}$. In the latter case two of the three linearly
independent state vectors are parasuperpartners.
\end{lem}

Finally we analyze the case $E>0$ and $|E,0\kt=0$, i.e., Case~(2). Again
following a similar argument as given in the previous case, one can show  that
$|E,\sqrt{2E}\kt\neq 0$ implies $|E,-\sqrt{2E}\kt\neq 0$ and vice versa.
Hence we can safely assume that in this case $E$ is doubly degenerate with the
eigenbasis $\{|E,\pm\sqrt{2E}\kt\}$.

Using Eq.~(\ref{e1.10}), one can still show the validity of (\ref{e3.1}).
However now Eq.~(\ref{e3.1}) can be written in the form:
	\be
	Q_2|E,\pm\sqrt{2E}\kt=b_{\pm}|E,\mp\sqrt{2E}\kt\;,
	\label{e5.2}
	\ee
where $b_\pm\in\C$. Furthermore, in view of (\ref{e5.2}) and (\ref{e12}), one
has
	\[0=(Q_2^3-2HQ_2)|E,\pm\sqrt{2E}\kt=-b_\pm(2E-b_+b_-)
	|E,\mp\sqrt{2E}\kt\;.\]
This means that either $b\pm=0$ or $b_+b_-=2E$. In fact, using Eq.~(\ref{e1.9})
one can easily show that the choice $b\pm=0$ leads to $E=0$ and is consequently
inadmissible. Therefore,
	\be
	 b_+b_-=2E\;.
	\label{e20}
	\ee
On the other hand, Eqs.~(\ref{e5.2}) and (\ref{e20}) can be employed to
compute:
	\[ |b_\pm|^2=\br E,\pm\sqrt{2E}|Q_2^2|E,\pm\sqrt{2E}\kt=
	\br E,\pm\sqrt{2E}|\lll Q_2^2|E,\pm\sqrt{2E}\kt\rrr=2E\:,\]
so that $b_\pm=\sqrt{2E}\exp(\pm i\lambda)$. Here Eq.~(\ref{e1.7}) is also used
and $\lambda\in\R$. Again it is possible to absorb the phase factors in
$|E,\pm\sqrt{2E}\kt$, in which case Eq.~(\ref{e5.2}) reads:
	\be
	Q_2|E,\pm\sqrt{2E}\kt=\sqrt{2E}|E,\mp\sqrt{2E}\kt\;.
	\label{e5.2'}
	\ee
The matrix representation of
$Q_1$, $Q_2$ and {\large$\tau$} is identical with the case of SQM
described in Sec.~2  (with $\kappa=2$). Thus the energy levels of this type
involve (up to subdegeneracies) a pair of parasuperpartners.

The following lemma summarizes the results of our analysis of Case~(2).
\begin{lem}
For $E>0$ and $|E,q_1=0\kt=0$, $E$ is doubly degenerate with the basis
$\{|E,\pm\sqrt{2E}\kt\}$. It consists of a pair of parasuperpartners.
\end{lem}

This concludes our treatment of the R-S  \psqm. Note that even in the presence
of other quantum numbers (subdegeneracies), the energy levels of type~(1b)
involve equal number of state vectors associated with the three basic
(subdegenerate) states. Similarly, the energy levels of type~(2) consist of an
equal number of state vectors of opposite chirality.  The latter is reminiscent
of the inclusion of SQM in \psqm \cite{rs}.


\section{Beckers-Debergh \psqm}

In terms of the self-adjoint parasupercharges (\ref{e7}), the B-D
parasuperalgebra (\ref{e4})--(\ref{e6}) is expressed by Eqs.~(\ref{e8}),
(\ref{e9}), (\ref{e10}) and
	\bea
	3\{Q_1,Q_2^2\}-2Q_1^3&=&2Q_1H
	\label{e7.1}\\
	3\{Q_2,Q_1^2\}-2Q_2^3&=&2Q_2H\;.
	\label{e7.2}
	\eea
Combining these relations with (\ref{e8}) and (\ref{e9}), one also has:
	\bea
	Q_1^3-3Q_2Q_1Q_2&=&2Q_1H
	\label{e7.3}\\
	Q_2^3-3Q_1Q_2Q_1&=&2Q_2H
	\label{e7.4}\\
	Q_1Q_2^2+Q_2^2Q_1-2Q_2Q_1Q_2&=&2Q_1H
	\label{e9.1}\\
	Q_2Q_1^2+Q_1^2Q_2-2Q_1Q_2Q_1&=&2Q_2H\:.
	\label{e9.2}
	\eea

Again one can check that (\ref{aa2}) with $\kappa=1/2$ is a special case of
(\ref{e7.1}) and
(\ref{e7.2}). Hence SQM is included in B-D \psqm as well.

In view  of Eqs.~(\ref{e10}), (\ref{e13}), (\ref{e14}),and (\ref{e7.3}), the
following is a straightforward  observation:
	\be
	Q_2Q_1Q_2|E,q_1\kt=\frac{q_1(q_1^2-2E)}{3}|E,q_1\kt\;.
	\label{e7.5}
	\ee
Next let us consider expressing Eq.~(\ref{e7.2}) in terms of its matrix
elements. A simple calculation yields:
	\be
	[(q_1-q_1')^2-2E]\br E,q_1'|Q_2|E,q_1\kt=0\;.
	\label{e8.0}
	\ee
Similarly, using (\ref{e8}) and (\ref{e7.5}), one has
	\be
	(q_1+q_1')\br  E,q_1'|Q_2^2|E,q_1\kt=
	\left[2q_1(q_1^2+E)/3\right]\delta_{q_1q_1'}\;.
	\label{e7.5.1}
	\ee
Setting $q_1=q_1'$ in the last equation, one finds:
	\be
	q_1\br E,q_1|Q_2^2|E,q_1\kt=q_1|Q_2|E,q_1\kt|^2=q_1(q_1^2+E)/3
	\;.
	\label{e7.7}
	\ee

Next, let us consider the negative energy levels. For $E<0$, Eq.~(\ref{e8.0})
requires $Q_2|E,q_1\kt=0$. This relation together with (\ref{e8}) imply
$q_1=0$. Thus, the negative energy levels are non-degenerate.

For the  $E=0$ energy level, Eq.~(\ref{e8.0}) implies
	\be
	{\rm either~~~}\br 0,q_1'|Q_2|0,q_1\kt=0\;,~~~~{\rm or}~~~q_1=q_1'\;.
	\label{e8.2}
	\ee
 In any case, $Q_2$ must be diagonal in the $E=0$ eigenspace:
	\be
	Q_2|0,q_1\kt=\kappa(q_1)|0,q_1\kt\;.
	\label{e21}
	\ee
Making use of this equation to simplify (\ref{e7.5.1}), one is led to:
	\be
	\kappa^2(q_1\neq 0)=q_1^2/3\;.
	\label{e22}
	\ee
Similarly, using (\ref{e21}) and (\ref{e7.4}), one obtains:
	\be
	\kappa(q_1)[\kappa(q_1)^2-3q_1^2]=0
	\label{e23}
	\ee
Eqs.~(\ref{e22}) and (\ref{e23}) imply $q_1=0$, and $\kappa(q_1=0)=0$. Hence,
$E=0$ also is non-degenerate.

The analysis of the positive energy levels is rather involved. We shall first
state some general algebraic results and then explore the following cases:
	\begin{itemize}
	\item[~] {\bf Case~(1)}:~$|E,0\kt\neq 0~~\mbox{and~~}\left\{
	\begin{array}{cc}
	\mbox{(1.a)~}&Q_2|E,0\kt\neq 0\\
	\mbox{(1.b)~}&Q_2|E,0\kt=0
	\end{array}\right.$
	\item[~] {\bf Case~(2)}:~$|E,0\kt=0$
	\end{itemize}

Let us suppose that for some $q_1\neq 0$, $|E,q_1\kt\neq 0$. then according to
Eq.~(\ref{e7.5.1}):
	\be
	Q_2^2|E,q_1\neq 0\kt=c_{q_1}|E,-q_1\kt+\frac{q_1^2+E}{3}|E,q_1\kt\;,
	\label{e9.7}
	\ee
where $c_{q_1}:=\br E,-q_1|Q_2^2|E,q_1\kt$ is a complex number. Note that
$|E,-q_1\kt$ may not exist, in which case one sets $|E,-q_1\kt=0$ and
$c_{q_1}=0$. It is possible that $|E,-q_1\kt\neq 0$ but still $c_{q_1}=0$.   We
shall next consider the case where $c_{q_1}=0$.

Let $q_1\neq 0$ and $c_{q_1}=0$, then Eqs.~(\ref{e9.7}) and (\ref{e7.4}) lead
to
	\be
	Q_1[Q_2|E,q_1\kt]=\frac{q_1^2-5E}{9q_1}[Q_2|E,q_1\kt]\;.
	\label{e9.8}
	\ee
Therefore by definition (\ref{e14}),
	\be
	Q_2|E,q_1\kt=\xi(q_1)|E,\tilde{q_1}\kt\;,~~~{\rm with~~~}
	\tilde{q_1}:=\frac{q_1^2-5E}{9q_1}\:,~\xi(q_1)\in\C\:.
	\label{e9.9}
	\ee
Next substitute (\ref{e9.9}) in (\ref{e9.7}). In view of $c_{q_1}=0$, this
yields:
	\[ \frac{q_1^3+E}{3}|E,q_1\kt=Q_2^2|E,q_1\kt=\xi(q_1)\xi(\tilde q_1)
	|E,\frac{\tilde q^2_1-5E}{9\tilde q_1}\kt\;,\]
and consequently:
	\be
	\frac{\tilde q^2_1-5E}{9\tilde q_1}=q_1\;,~~{\rm so~that~~}
	q_1=\pm\sqrt{E/2}\;.
	\label{e10.1}
	\ee
Moreover, in view of (\ref{e9.7}), (\ref{e9.9}) and (\ref{e10.1}),
$\xi(q_1)=\sqrt{E/2}\: e^{i\gamma}$. Again redefining the phases of
$|E,\pm\sqrt{E/2}\kt$, one is able to set $\gamma=0$. The result is
	\bea
	Q_2|E\pm\sqrt{E/2}\kt&=&\sqrt{E/2}~|E,\mp\sqrt{E/2}\kt.
	\label{e10.2}\\
	Q_2^2|E,\pm\sqrt{E/2}\kt&=&(E/2)~|E,\pm\sqrt{E/2}\kt\;.
	\label{e10.2'}
	\eea
Note that the last two relations also imply
	\be
	|E,\sqrt{E/2}\kt=0~~~~{\rm if~and~only~if}~~~~|E,-\sqrt{E/2}\kt=0\;.
	\label{e25}
	\ee

Next consider the case $q_1\neq 0$ and $c_{q_1}\neq 0$. This means that
necessarily $|E,\pm q_1\kt\neq 0$. Let both sides of (\ref{e7.4}) act on
$|E,q_1\kt$ from the left and use (\ref{e9.7}) to compute $Q_2^3|E,q_1\kt$.
This leads to
	\[ Q_1Q_2|E,q_1\neq 0\kt=\frac{1}{3q_1}\left[
	(\frac{q^2_1+E}{3}-2E)Q_2|E,q_1\kt+c_{q_1}Q_2|E,-q_1\kt\right]\;.\]
Now multiplying  both sides of this expression by $Q_2$ from the left and using
Eqs.~(\ref{e7.5}) and (\ref{e9.7}) to simplify the result, one arrive at:
	\be
	\left[(8q_1^4-14q_1^2E+5E^2-9c_{q_1}c_{-q_1})/3\right]~|E,q\kt+
	2c_{q_1}(2E-q_1^2)~|E,-q_1\kt=0\;.
	\label{e30}
	\ee
Therefore,
	\bea
	q_1&=&\pm\sqrt{2E}
	\label{e31}\\
	c_{q_1}c_{-q_1}&=&|c_{q_1}|^2\: =\:E^2\;,
	\label{e32}
	\eea
where we also have  used the obvious identity $c_{-q_1}=c_{q_1}^*$.
At this stage we begin considering all possible cases.

First suppose that $E>0$ and $|E,0\kt\neq 0$,  (Case~(1)). Then acting both
sides of (\ref{e7.1}) on $|E,0\kt$ from the left, one finds
$Q_1Q_2^2|E,0\kt=0$. Thus, by definition (\ref{e14}),
$Q_2^2|E,0\kt=\alpha|E,0\kt$. Let us further assume that $Q_2|E,0\kt\neq 0$,
i.e., consider Case~(1.a). Then the coefficient $\alpha$ can be determined by
Eq.~(\ref{e7.4}). The result is $\alpha=2E$, so that
	\be
	Q_2^2|E,0\kt=2E|E,0\kt\;.
	\label{e11.1}
	\ee
Moreover, making use of (\ref{e7.2}), one also finds:
	\be
	Q_1^2[Q_2|E,0\kt]=2E[Q_2|E,0\kt]\;.
	\label{e11.5.1}
	\ee
Hence clearly
	\be
	\br E,0|Q_2|E,0\kt=0\;.
	\label{e11.5.2}
	\ee
Since by the hypothesis $Q_2|E,0\kt\neq 0$, there must be some $q_1\neq 0$,
such that $\br E,q_1|Q_2|E,0\kt\neq 0$. Multiplying $\br E,q_1|$ by both sides
of (\ref{e11.5.1}), one immediately finds
	\be
	q_1=\pm\sqrt{2E}\;.
	\label{e11.5.2'}
	\ee
Hence in view of Eq.~(\ref{e11.5.2}),
	\be
	Q_2|E,0\kt=\tilde a_+|E,\sqrt{2E}\kt+\tilde a_-|E,-\sqrt{2E}\kt\;.
	\label{e11.7}
	\ee

Using the last equation and Eq.~(\ref{e7.5}), i.e., $Q_2Q_1Q_2|E,0\kt=0$,
one can further show
	\be
	\tilde a_+Q_2|E,\sqrt{2E}\kt=\tilde a_-Q_2|E,-\sqrt{2E}\kt\;.
	\label{e11.8}
	\ee
This in turn implies that the vanishing of either of $|E,\pm\sqrt{2E}\kt$
implies the vanishing of $Q_2^2|E,0\kt$ and consequently of $Q_2|E,0\kt$.
This is inconsistent with the hypothesis of this case. Thus
$|E,\pm\sqrt{2E}\kt\neq 0$.

A remarkable consistency check on our analysis is to observe the coincidence of
Eqs.~(\ref{e11.5.2'}) and (\ref{e31}). In particular, $c_{\pm\sqrt{2E}}\neq 0$.

To obtain the action of $Q_2$ on eigenstates of $E$, we proceed as follows.
Since, $Q_2^2|E,0\kt\neq 0$, Eqs.~(\ref{e11.7}) and (\ref{e11.8}) imply
$Q_2|E,\pm\sqrt{2E}\kt\neq 0$. in fact, substituting (\ref{e11.7}) in
(\ref{e11.5.2}) and using (\ref{e11.8}), one has:
	\bea
	\tilde  Q_2|E,\pm\sqrt{2E}\kt&=&\sqrt{E}\,e^{i\tilde\gamma_\pm}\,|E,0\kt
	\label{e11.9'}\\
	Q_2|E,0\kt&=&\sqrt{E}\lll e^{-i\tilde\gamma_+}|E,\sqrt{2E}\kt+
	e^{-i\tilde\gamma_-}|E,-\sqrt{2E}\kt\rrr\;,
	\label{e11.7'}
	\eea
where $e^{i\tilde\gamma_\pm}:=\sqrt{E}/\tilde a_\pm$ are unimportant phase
factors.
A remarkable observation is that Eqs.~(\ref{e11.9'}) and (\ref{e11.7'}) are
identical with Eqs.~(\ref{e18'}) and (\ref{w0'}) with $\zeta=1$. Thus the
analysis of Case~(1.b) of the previous section -- with $\zeta=1$ -- applies for
the case considered here.

A summary of the analysis of Case~(1.a) is given by
\begin{lem}
For $E>0$, $|E,0\kt\neq 0$, and $Q_2|E,0\kt\neq 0$, $E$ is triply degenerate
with the eigenbasis $\{|E,0\kt,|E,\pm\sqrt{2E}\kt\}$.
\end{lem}

Next we consider Case~(1.b), where $E>0,~|E,0\kt\neq 0$ and $Q_2|E,0\kt=0$.
In this case $E$ may either be non-degenerate with $|E,0\kt$ representing  the
non-degenerate state vector, or it may be degenerate. In the latter case, there
must be some $q_1\neq 0$ with $|E,q_1\kt\neq 0$. Suppose that $|E,-q_1\kt=0$.
Then, $c_{q_1}$ of Eq.~(\ref{e9.7}) must vanish. But in this case, according to
Eqs.~(\ref{e10.1}) and (\ref{e25}), $q_1=\pm\sqrt{E/2}$ and $|E,-q_1\kt=0$
implies
$|E,q_1\kt=0$. This is a contradiction. Hence, $|E,\pm q_1\kt\neq 0$.

In fact, one can show that indeed the condition $Q_2|E,0\kt=0$ implies
$c_{q_1}=0$. To see this, we  employ Eq.~(\ref{e8.0}) which states that
	\[Q_2|E,q_1\kt=\mu |E,0\kt+\nu |E,-q_1\kt\;.\]
Then,
	\[c_{q_1}:=\br E,-q_1|\lll Q_2^2|E,q_1\kt\rrr=
	\nu\br E,-q_1|Q_2|E,-q_1\kt=0\;,\]
where we have used $Q_2|E,0\kt=0$ and Eq.~(\ref{e8.0}). In view of this
observation we conclude:
\begin{lem}
For $E>0,~|E,0\kt\neq 0$ and $Q_2|E,0\kt=0$, $E$ is either non-degenerate or it
is triply degenerate with the basis $\{ |E,0\kt,|E,\pm\sqrt{E/2}\kt\}$.
\end{lem}
Furthermore the basic eigenvectors are related via Eq.~(\ref{e10.2}). This case
is quite similar to the case (1.b.ii) of the previous section. In fact, our
treatment of Case~(1b) of the R-S \psqm with $\zeta=0$ applies to Case~(1.b) of
B-D \psqm. The only difference is that in the latter case we need to set
$\zeta=0$ and multiply the right hand side of  Eqs.~(\ref{rep}) and
(\ref{tau5}) by 1/2.

This leaves us with Case~(2), where $E>0$ and $|E,0\kt=0$. In this case,
$Q_2|E,0\kt=0$ is trivially satisfied and the analysis of the previous case
applies. In particular:
\begin{lem}
For $E>0$ and $|E,0\kt=0$, $E$ is doubly degenerate with the basic state
vectors, $|E,\pm\sqrt{E/2}\kt$, being parasuperpartners.
\end{lem}
This case is identical with the case of SQM (with  $\kappa=1/2$). This is a
clear indication of the inclusion of SQM in B-D \psqm.

This concludes our analysis of B-D \psqm.


\section{Parasupersymmetric Topological Invariants}

In the previous sections it was shown that unlike SQM, the degeneracy structure
of the \psqm does not allow the Witten index (:=trace({\large$\tau$})) or
similar quantities to be invariant under the smooth deformations of the
Hamiltonian.
There are three obvious reasons justifying this remark. These are:
\begin{itemize}
\item[~]{\bf a)}  existence of negative energy levels;
\item[~]{\bf b)}  existence of positive energy levels with $\br Q_1\kt_E=\br
Q_2\kt_E=0$ (these are referred to as non-degenerate levels in Secs.~3 and 4);
\item[~]{\bf c)}  existence of both doubly and triply degenerate energy levels.
\end{itemize}

In order  to find systems with invariants analogous to trace({\bf $\tau$}) of
SQM, one needs to find \psqm systems for which none of the above obstacles
occur. In this section we consider an example of a class of \psqm systems which
fulfill this requirement and therewith lead to  ``new" topological
invariants.\footnote{The identification of such invariants with known or
unknown topological invariants is the subject of further investigation.}

Consider a R-S $(p=2)$--PSQ Mechanical system whose Hamiltonian is given by
	\be
	H=\frac{\gamma}{2} \left[ (\qq\qq^\dagger)^2 +
	(\qq^\dagger\qq)^2+\alpha(\qq\qq^{\dagger 2}\qq +  \qq^\dagger
	\qq^2\qq^\dagger)\right]^{\frac{1}{2}}\;,
	\label{k1}
	\ee
where $\alpha$ and $\gamma >0$ are real parameter. The system (\ref{k1}) with
$\alpha=-1/2$, $\gamma=1$, and
$\qq\qq^{\dagger 2}\qq=\qq^\dagger\qq^2\qq^\dagger$ has been suggested by Khare
\cite{kh3} and investigated in the context of ordinary one-dimensional quantum
mechanics for arbitrary $p$. Here we shall assume that $p=2$ and that the
system also satisfies the parasuperalgebra of R-S (\ref{e1})--(\ref{e3}).

In terms of the self-adjoint parasupercharges (\ref{e7}), Eq.~(\ref{k1}) is
written in the form:
	\be
	H=\frac{\gamma}{2}\left[ \lll\frac{1+\alpha}{2}\rrr (Q_1^2+Q_2^2)^2
	-\lll \frac{1-\alpha}{2}\rrr ([Q_1,Q_2])^2\right]^{\frac{1}{2}}\;.
	\label{k3}
	\ee
Clearly, the case $\alpha=1$ and $\gamma=1/2$  corresponds to SQM with
$\kappa=2$.

In addition, the negative and ``non-degenerate" positive energy levels are now
forbidden. To see this it is sufficient to square both sides of (\ref{k3}),
i.e., consider
	\be
	4 H^2=\gamma^2\left[ \lll\frac{1+\alpha}{2}\rrr (Q_1^2+Q_2^2)^2
	-\lll \frac{1-\alpha}{2}\rrr ([Q_1,Q_2])^2\right]\;,
	\label{k4}
	\ee
and note that according to the results of Sec.~3, all such states are
eliminated by $Q_2$. Thus upon the action of both sides of (\ref{k4}) on such
states,
the right hand side vanishes whereas the left hand side does not.

Furthermore, it can be shown quite easily that the doubly degenerate levels
of lemma~2 are also missing for all values of $\gamma$ except $1/2$. To see
this, we use the representation of $Q_1$ and $Q_2$  to express the right hand
side of (\ref{k4}). The left hand
side equals $4E^2$ times the $2\times 2$ unit matrix. A simple calculation
proves our assertion, i.e., for $\gamma\neq 1/2$, Eq.~(\ref{k4}) is not
satisfied. Thus  the energy levels cannot be doubly degenerate.

This leaves us only with the triply degenerate states. We must still check that
for these states, Eq.~(\ref{k4}) can indeed be satisfied. Otherwise
Eq.~(\ref{k1}) would be inconsistent with the R-S parasuperalgebra
(\ref{e1})--(\ref{e3}). Khare \cite{kh3} has shown that for $\alpha=-1/2$,
there are examples for which both (\ref{k1}) and (\ref{e1})--(\ref{e3}) are
fulfilled. In fact, as we shall demonstrate instantly,  the condition
$\alpha=-1/2$ also is  necessary.

Substituting Eq.~(\ref{rep}) in the right hand side of (\ref{k4}) and carrying
out the calculations, one observes that Eq.~(\ref{k4}) is satisfied
only if $\alpha=-1/2$, $\gamma=1$, and $\zeta=1$. In this case,  according to
(\ref{rep}) and (\ref{tau6}), the parasupercharges are represented by
	\be
	\left. Q_1\right|_{{\cal H}_E}=\sqrt{2E}J_3^{~(1)}\;,
	\left. Q_2\right|_{{\cal H}_E}=\sqrt{2E}J_1^{~(1)}\;,~~
	\forall E>0\;.
	\label{k5}
	\ee
Note the remarkable similarity between  Eqs.~(\ref{k5}) and (\ref{x2}).

Summarizing the results, one has
\begin{lem}
For the R-S \psqm the condition that the Hamiltonian satisfies (\ref{k1})
implies that
the energy eigenvalues are non-negative and that  either $\gamma=1/2$ and
$\alpha$ is
arbitrary, or $\gamma=1$ and $\alpha=-1/2$. In the former case,  the positive
energy levels
consist of (para)superpartners. In the latter case
the positive energy levels are triply degenerate with two of the eigenstates
being parasuperpartners.
\end{lem}

In fact, one can similarly show that a B-D parasupersymmetric system which
satisfies (\ref{k1}) has precisely identical degeneracy structure. In this
case, either $\gamma=2$ and positive energy levels are doubly degenerate, or
$\gamma=1$ and $\alpha=-1/2$, in which case the positive energy levels are
triply degenerate.

Now let us consider smooth deformations of the (parameters of the) Hamiltonian
satisfying the R-S parasuperalgebra and  Eq.~(\ref{k1}) with
$\alpha=-\gamma/2=-1/2$. Then according to   Lemma~6, the initial zero-energy
states can only acquire  positive energy in groups of three and vice versa a
positive energy state can only collapse to the zero level if the other two
states within the original (non-perturbed) level accompany it. Thus if the
chirality operator {\large$\tau$} has the same
signature, say det({\large$\left.\tau\right|_{{\cal H}_E}$})$=-1$, for all
$E>0$ then the quantity
	\be
	\Delta_{(p=2)}:= n^{(\pi B)}-2n^{(\pi F)}= n_0^{(\pi
	B)}-2n_0^{(\pi F)}\;,
	\label{k6}
	\ee
with
	\bea
	n^{(\pi B)}&:=&\mbox{ number of parabosonic states;}\nn\\
	n^{(\pi F)}&:=&\mbox{ number of parafermionic states;}\nn\\
	n_0^{(\pi B)}&:=&\mbox{ number of parabosonic states of zero
	energy;}\nn\\
	n_0^{(\pi B)}&:=&\mbox{ number of parabosonic states of zero
	 energy,}\nn\\
	\eea
remains invariant under the deformation. $\Delta_{(p=2)}$ is a generalization
of the Witten index of SQM and in the above sense is a topological invariant.
In fact, if the corresponding system is defined on a smooth manifold or a fiber
bundle, i.e., the Hamiltonian and parasupercharges depend on the corresponding
geometric structures (connection), then $\Delta_{(p=2)}$ is a true topological
invariant.

By a similar argument one can also show that if a R-S (B-D) system
satisfies Eq.~(\ref{k1}) with $\gamma=1/2$ (resp.  $\gamma=2$), then the
	\[ {\rm Witten~ index} :=\mbox{ trace({\large$\tau$})}=n^{(\pi B)}-n^{(\pi
F)}= n_0^{(\pi
	B)}-n_0^{(\pi F)}\;,\]
is a topological invariant.

The topological invariants defined in this way are measures of the exactness of
 the parasupersymmetry. More precisely,  their being non-zero implies the
existence of
zero-energy ground states and consequently the exactness of parasupersymmetry.


\section{Conclusion}

The Robakov--Spiridonov and Beckers--Debergh ($p=2$) parasupersymmetric quantum
systems share identical general degeneracy structures.  For the R-S \psqm, we
defined a continuous ($\zeta$) and a discrete parameter ($\epsilon$) for each
triply degenerate energy level. These
determine the action of $Q_2$ on the corresponding  states. The triply
degenerate energy levels of the B-D \psqm have the same structure as those of
the R-S systems with $\zeta=0$ or $\zeta=1$. Thus in this sense, the R-S \psqm
is more general than the B-D \psqm.
The results of the present article is consistent with the results obtained for
specific examples
considered in the literature \cite{rs,bd}.

A possible direction of further investigation is  to seek   a physical
interpretation for the parameters $\zeta$ and $\epsilon$ of the  R-S \psqm.
Another direction is to try to employ similar methods to arbitrary
$(p>2)$--parasupersymmetry. The parasuperalgebra for general R-S
($p>2)$--parasupersymmetry is considerably more complicated. But its variations
\cite{kh2} may be attacked by similar methods.  As demonstrated for
one-dimensional systems by Khare \cite{kh2},
the degeneracy structure of the systems with even $p$  (respectively odd $p$)
displays similar features. This makes the study of $(p=odd)$ parasupersymmetry
more interesting.  In fact, one might hope that they  involve   similar or even
 more appealing  phenomena than supersymmetric quantum mechanics.

The treatment of the \psqm presented in this article uses only the defining
parasuperalgebras.
The only additional assumption made here is that the state vectors belong to a
Hilbert space.
Therefore, the results obtained here are applicable to arbitrary quantum
systems satisfying these parasuperalgebras. In particular, one might consider
systems which are sensitive to the geometric structure of an arbitrary manifold
or a fiber bundle.

We have also  introduced a chirality (parasupersymmetric involution) operator
{\large$\tau$},
and  studied its possible representations for different types of energy levels.
The general degeneracy structure of the \psqm does not allow for  the Witten
index, i.e.,  trace({\large$\tau$}), to be a topological invariant.  The main
obstacle  is the possible existence of negative and positive ``non-degenerate''
 ($\br Q_1\kt_E=\br Q_2\kt_E=0$) energy levels and the fact that even the
degenerate levels can be either two or three fold degenerate.

These obstacles do not survive if one considers R-S or B-D type
parasupersymmetric systems whose Hamiltonian are given by Eq.~(\ref{k1}).
For such systems,   all the positive energy levels are either  two fold or
three fold degenerate.
For the systems with doubly degenerate positive energy levels the Witten index
remains to be a topological invariant. For the systems with triply degenerate
positive energy levels,  the difference of the number of parabosonic states and
twice of the number of parafermionic states is a topological invariant.  In
both cases the non-vanishing of the corresponding topological invariant is an
indication of the exactness of the parasupersymmetry.
The study of  specific examples of these  systems  and the identification of
the topological invariants introduced above is the subject of ongoing
investigation.

.
\end{document}